\documentclass[twocolumn]{aastex61}

\begin{document}

\title{effect of massive neutrinos on the position of cold dark matter halo: revealed via Delaunay Triangulation void}

\correspondingauthor{Tong-Jie Zhang}
\email{tjzhang@bnu.edu.cn}

\author{Jian Qin}
\affiliation{Department of Astronomy,Beijing Normal University,Beijing 100875,China}

\author{Yu Liang}
\affiliation{Tsinghua Center for Astrophysics (THCA) \& Department of Physics, Tsinghua University, Beijing 100084, China }

\author{Cheng Zhao}
\affiliation{Tsinghua Center for Astrophysics (THCA) \& Department of Physics, Tsinghua University, Beijing 100084, China }

\author{Hao-Ran Yu}
\affiliation{Tsung-Dao Lee Institute, Shanghai Jiao Tong University, Shanghai, 200240, China}
\affiliation{Canadian Institute for Theoretical Astrophysics, University of Toronto, Toronto, ON M5H 3H8 Canada}

\author{Yu Liu}
\affiliation{Department of Astronomy,Beijing Normal University,Beijing 100875,China}

\author{Tong-Jie Zhang}
\affiliation{Department of Astronomy,Beijing Normal University,Beijing 100875,China}
\affiliation{Shandong Provincial Key Laboratory of Biophysics, School of Physics and Electric Information, Dezhou University, Dezhou 253023, China}

\begin{abstract}
Using cosmological $N$-body simulation which coevolves cold dark matter (CDM) and neutrino particles, we discover the local effect of massive neutrinos on the spatial distribution of CDM halos, reflected on properties of the Delaunay Triangulation (DT) voids.
Smaller voids are generally in regions with higher neutrino abundance and so their surrounding halos are impacted by a stronger neutrino free streaming. This makes the voids larger (surrounding halos being washed outward the void center). On the contrary, larger voids are generally in regions with lower neutrino abundance and so their surrounding halos are less impacted by neutrino free streaming, making the voids smaller (surrounding halos being squeezed toward the void center). 
This characteristic change of the spatial distribution of the halos suppresses the 2-point correlation function of halos on scales $\sim$ 1 Mpc$/h$ and significantly skews the number function of the DT voids, which serve as measurable neutrino effects in current or future galaxy surveys.
\par
\end{abstract}


\section{Introduction} \label{sec:intro}

Neutrino is one of the most fundamental particles in the Universe. 
Flavour oscillation experiments discover that at least two types of neutrinos have non-zero mass \citep{1998PhRvL..81.1562F,2001PhRvL..87g1301A,2002PhRvL..89a1301A} and yield a lower limit on the sum of their mass: $M_{\nu}\equiv \sum _{i=1} ^3 m_{\nu _i} \gtrsim 0.05$ \citep{2014ChPhC..38i0001O}.  From a cosmological point of view, the neutrino masses modulate the matter-to-radiation ratio which leaves an imprint on the Cosmic Microwave Background (CMB) and their high thermal velocities suppress the matter power spectrum on the small scales. Based on this, the 
sum of the neutrino masses $M_{\nu}$ can be inferred from cosmological observations. For example, the recent CMB measurement provides a upper limit of $M_{\nu} \lesssim 0.23$ eV \citep{2016A&A...594A..13P} and combined the large-scale structure (LSS) surveys this limit is placed to be 0.15 eV \citep{2017MNRAS.470.2617A}.
The upcoming cosmological large-scale structure (LSS) surveys such as  Euclid \citep{2011arXiv1110.3193L}, LSST \citep{2009arXiv0912.0201L} and eBOSS \citep{2016AJ....151...44D} are expected to improve this limit.

While the effects of massive neutrinos on the CMB and on the LSS in the early Universe are well described by the linear perturbation theory \citep{2006PhR...429..307L,2013neco.book.....L}, their impact on the LSS at late time become fully non-linear. This arouses the need for cosmological $N$-body simulations that coevolve cold dark matter (CDM) and neutrino particles, by which we are able to study the non-linear effects of massive neutrinos on LSS reflected on, for example, the halo mass function \citep{2010JCAP...09..014B}, the galaxy bias \citep{2014JCAP...02..049C}, and the properties of cosmic voids \citep{2015JCAP...11..018M}. 
Both the cosmic variance and the neutrino Poisson noise limit the presition of measuring the neutrino effects from $N$-body simulations.  These two noises can be controlled by choosing a bigger simulation volume and a larger neutrino particle number density, which is computationally challenging.
 
In this work by using one of the world's largest $N$-body simulations, which coevolves 3 trillion CDM and neutrino particles in a box of 1200 Mpc$/h$, we are able to precisely investigate the non-linear effect of massive neutrinos. 
This simulation named ``TianNu'' is run on Tianhe-2 supercomputer and has been used to precisely measure the cold dark matter-neutrino dipole \citep{2016arXiv161009354I}.
And more recently it has been used to study the differential neutrino condensation, which discovers that the CDM halo masses are altered according to different neutrino environments \citep{2017NatAs...1E.143Y}.
Following this idea we focus our work on investigating how the positions of the CDM halos are altered in regions of different neutrino abundance.   
However in order to reveal the local changes of the halo positions of different regions, it is not suitable to use a global statistics such as the power spectrum or the two-point correlation function.
Alternatively, the Delaunay trIangulation Void findEr (DIVE) \citep{2016MNRAS.459.2670Z} based on  Delaunay trIangulation (DT) constructs voids geometrically from the spatial distribution of halos and divides the Universe into different spherical regions with radius ranging from $\sim$1 Mpc$/h$ to $\sim$35 Mpc$/h$. These spherical regions can serve as tracers for regions of different neutrino environment as well as probes for detecting the local changes of the halo positions in these regions. 

This paper is organized as follows. In Section \ref{sec:methods}, we utilize the DIVE DT void finder to construct void catalogs from TianNu and TianZero simulations and utilize the simulation outputs to measure the environment neutrino density. In Section \ref{sec:results}, choosing the void radius difference between TianNu and TianZero as the proxies of the local changes of the halo positions we present the void radius changes in different neutrino environment as well as the suppressed halo 2-point correlation function and the skewed void number function. We summarize and discuss the results in Section \ref{sec:Summary}.  

\section{methods} \label{sec:methods}
\subsection{TianNu Simulation} \label{sec:simulation}
TianNu imposes a flat Universe with cosmological parameters [$\Omega_c,\Omega_b,h,n_s,\sigma_8$] = [0.27,0.05,0.67,0.96,0.83],
where $\Omega_c,\Omega_b,h,n_s$ and $\sigma_8$ are respectively
the densities of CDM and baryons, Hubble's parameter,
the initial tilt and fluctuations of the power spectrum.
The choice of neutrino mass models the minimal `normal hierarchy' with two light species included in the background cosmology and one heavy species ($m_{\nu}=0.05 eV$) that are traced with $N$-body particles. 
The simulation follows the evolution of 6912$^3$ CDM particles and 13824$^3$ neutrino particles in a periodic box of size 1200 Mpc$/h$.
The comparison neutrino-free simulation ``TianZero'' is given the same CDM initial conditions and $\Omega_m$ (total matter density) as TianNu and is equivalent to a TianNu simulation in which $\sum m_{\nu}$=0. The halos are obtained using a spherical overdensity approach and we regard the center-of-mass of each halo as its position. Details about the simulations and the halo finder are introduced in \cite{2017NatAs...1E.143Y}.

\begin{figure}[ht!]    
\center{\includegraphics[width=9cm]  {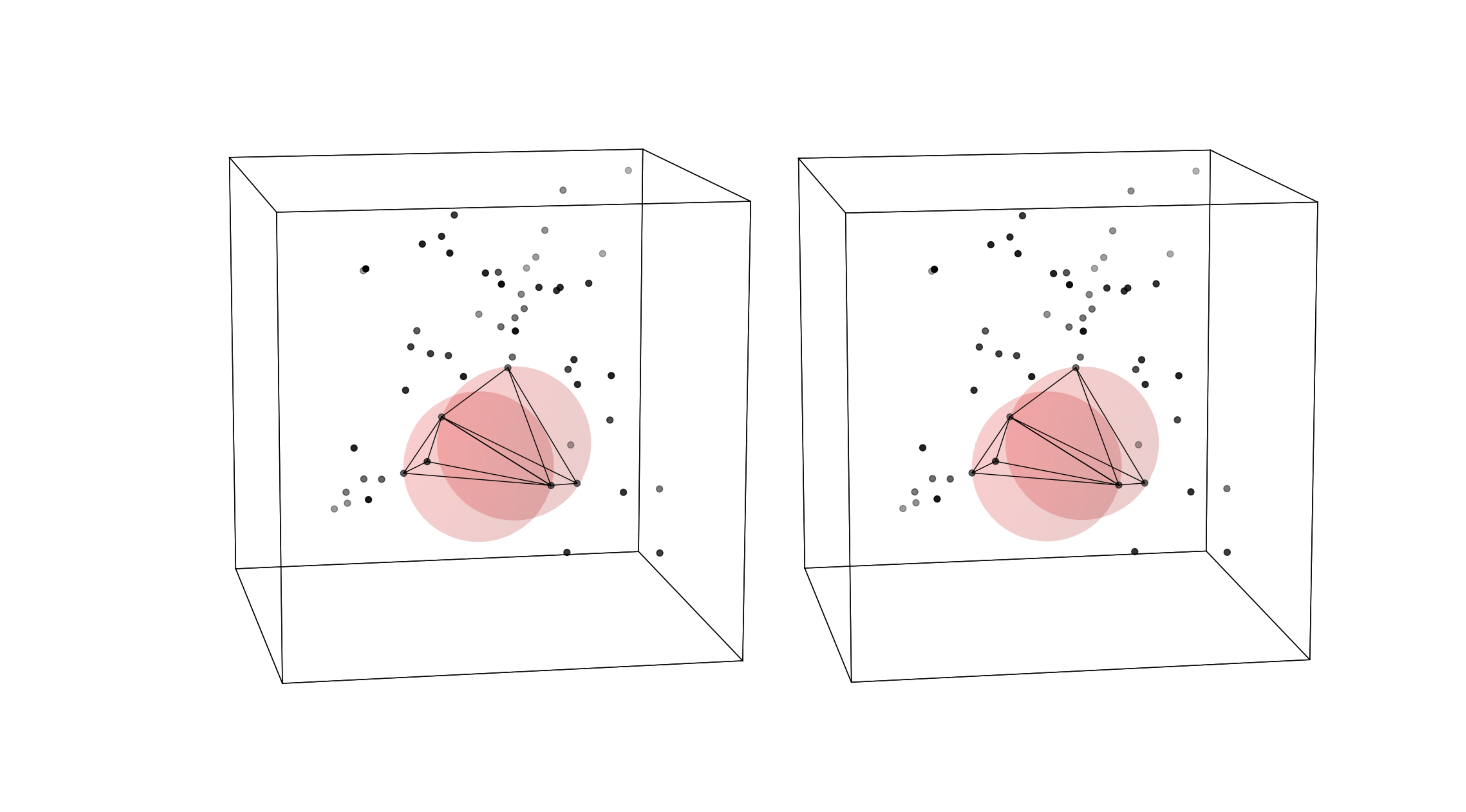}}        
\caption{The halos (black points) from the same parts of the TianZero (left) and TianNu (right) simulations at redshift $z=0.01$, respectively in a box with $50^3$ Mpc$^3 h^{-3}$ volume and with halo number density $\sim 3.5\times 10^{-4} $Mpc$^{-3}h^3$. 
The red spheres denote part of the voids recognized by the DIVE technique. The halo position variances between TianNu and TianZero simulations are very small and so are the variances of the void positions and radius, making it feasible to match halo and void pairs between the two simulations.  
}
\label{3D_intuitive}
\end{figure}

The comparison of the halo spatial distributions between TianNu and TianZero is illustrated in Figure \ref{3D_intuitive}. 
This figure shows a high but not surprising similarity of the halos (the black solid points) between the two simulations because the effect of the massive neutrinos is exactly very small. 
This similarity makes it feasible to associate the same halos between the two simulations, so the halo distribution change between the two simulations can be locally detected through measuring the changes of the DT voids (denoted with the red spheres in Figure \ref{3D_intuitive}).  
Quantitatively we pick a halo pair between TianNu and TianZero if two halos 
are separated within 100 kpc/h 
in their respective volumes and their mass vary by less than 10\%.
Each halo pair represents the
evolution of the same physical object in different cosmologies. 
We drop a pair when one of the halos of the pair can match multiple halos from another simulation, corresponding to the case that two different halos with nearly same masses are located very close.
And we can associate $\sim$94\% of all halos of TianNu and TianZero in this way. 
The code details and its scale to Tianhe-2 supercomputer are introduced in \cite{2017RAA....17...85E}. Analysis and results in this paper are based on the checkpoint of the two simulations at redshift $z=0.01$. 

\subsection{DT Void} \label{subsec:DIVE}
\begin{figure}[ht!]   
\center{\includegraphics[width=6cm]  {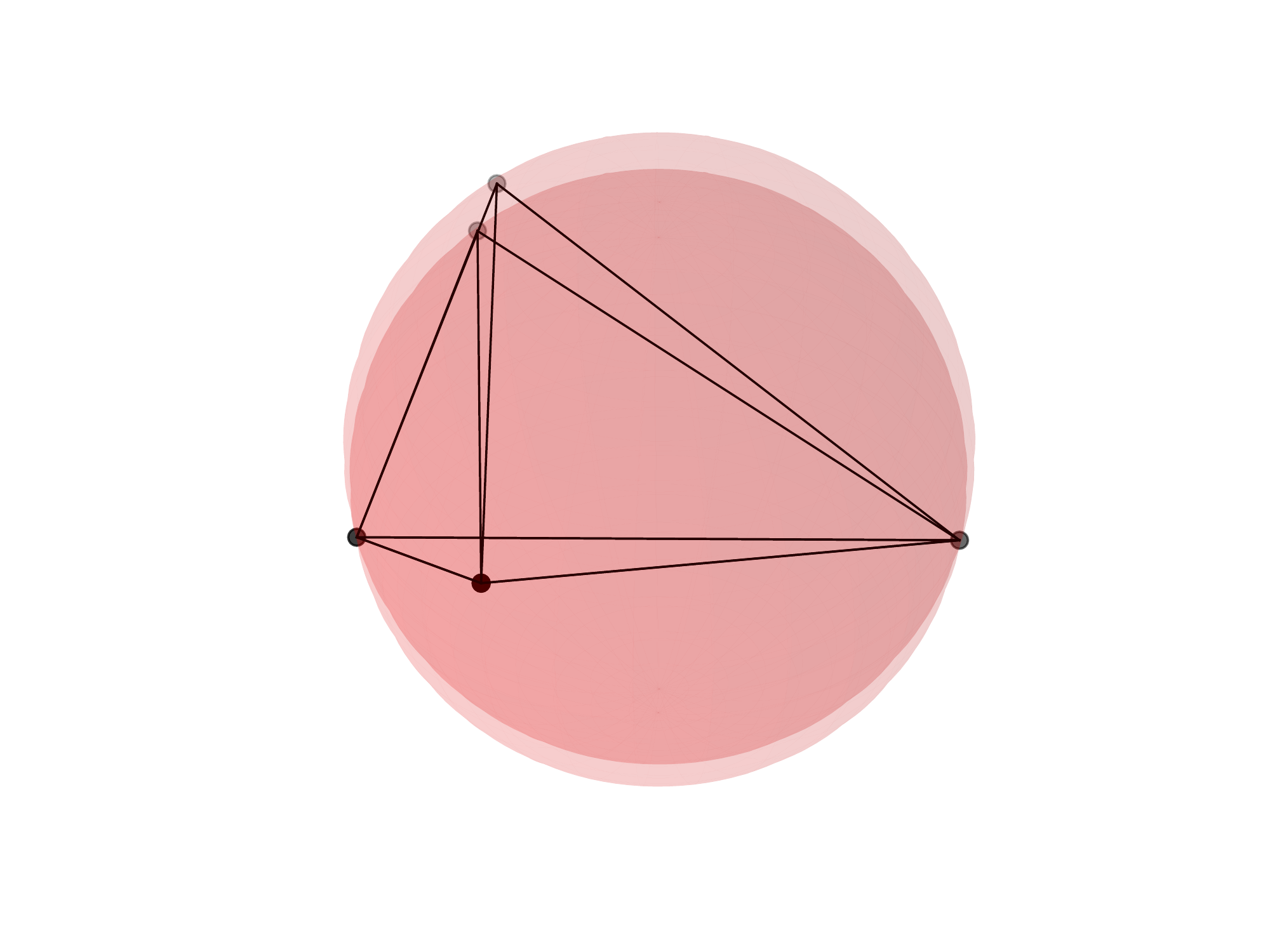}}        
\caption{Intuitive display of the changes of the DT void  (the red spheres) caused by the motion of the halo. The constraining halos are denoted with black solid points and we have artificially moved the halo 1.0 Mpc$/h$ in order to have a clear demonstration. Void will be larger (smaller) if the constraining halos move outward (toward) the void center. Comparing TianNu and TianZero simulations, the actual halo position variances between the halo pairs are of order 10 kpc$/h$ and the void radius variances are less than 200 kpc$/h$.
\label{2D_intuitive} } 
\end{figure}

DIVE DT void finder is based on Delaunay Triangulation (DT) \citep{2016MNRAS.459.2670Z} and has been used to measure BAO signals for the first time from the clustering of voids both theoretically \citep{2016MNRAS.459.4020L} and observationally \citep{2016PhRvL.116q1301K}.
Circumsphere constrained by tetrahedron of halos is defined as DT void 
if it does not contain any halos within its radius. 
Therefore DT void catalog depend entirely on the halo spatial distribution, and the position changes of the halos lead to the changes of the voids, intuitively displayed in Figure \ref{2D_intuitive}. Generally, halos moving toward the void center results in a reduced void radius while halos moving outward leads to an increased void radius. 
We construct DT void catalogs of TianNu and TianZero from the most massive $\sim$0.42, $\sim$0.59 and $\sim$0.75 million halo pairs corresponding to a halo number density of $\{2.5,3.5,4.5\}\times 10^{-4} $Mpc$^{-3}h^3$ respectively, and regard the geometric (spherical) center of each void as its position. We match void pairs in a similar way of matching halo pairs by defining a void pair between TianNu and TianZero if the two voids are separated within 100 kpc/h in their respective volumes and their radii vary by less than 1\%.
The difference in each void pair represents the position differences of the same constraining halos in different cosmologies (indicated with the vertexes of the tetrahedrons in Figure \ref{3D_intuitive} \& \ref{2D_intuitive}). Particularly, the radius difference of the voids indicates that the constraining halos are moving toward or outward of the void centers.
A pair is dropped if one of the voids of the pair can match multiple voids from another simulation, referring to the case that two different voids are closely overlapped to each other. We can associate $\sim$86\% of all voids between TianNu and TianZero in this way.

\subsection{Neutrino Environment}
We estimate the envirenment neutrino density contrasts 
$\delta_\nu$ centered on the position of each void, where $\delta_\nu$ on scale $k^{-1}\simeq 60$ ${\rm Mpc}/h$ are obtained 
from the neutrino density fields $\rho$ from the simulation outputs with
\begin{figure*}[ht!]
\includegraphics[width=18cm]{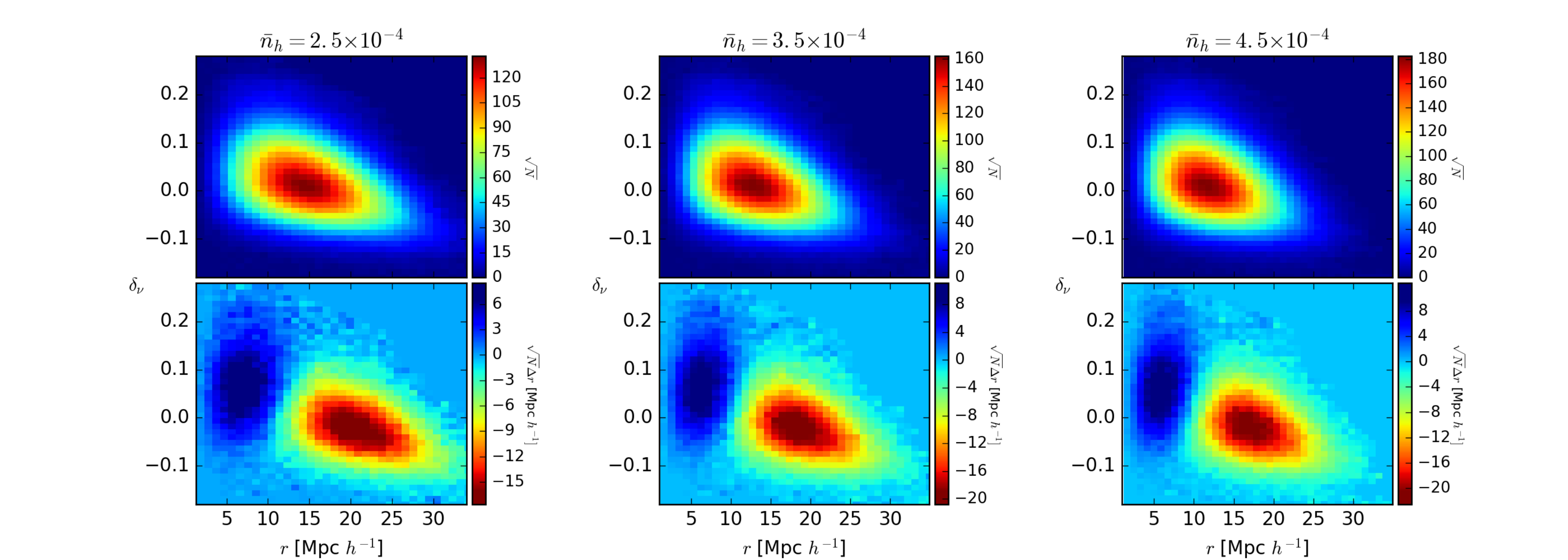}
\caption{(Upper panels) Distribution of voids as a
  function of $\delta _\nu$ and $r$.
  This exhibits the anti-correlation between the two
  quantities: voids of small (large) radius have generally high (low) $\delta _\nu$. (Lower panels) Variation in void radius 
caused by massive neutrinos.
The color scale shows a weighted histogram of the 
  radius variations observed between TianNu and
  TianZero, $\Delta r \equiv r_\nu-r_0$, organized
  in bins of void radius and neutrino density contrast. 
The voids are divided into two clusters in the $\delta _\nu$-$r$ plane. For smaller voids affected by a relatively stronger neutrino free streaming (larger $\delta _\nu$), their surrounding halos are washed outward the void centers, increasing the void radius (the bluer cluster). Conversely, for larger voids affected by a weaker (smaller $\delta _\nu$), their surrounding halos are squeezed toward the void centers, decreasing the void radius (the redder cluster).
The three columns show the results with voids constructed from halo catalogs of different halo number densities $\bar{n}_h$, in units of $h^3\rm{Mpc}^{-3}$.
\label{r_e_v} }     
\end{figure*}
$\delta_\nu \equiv \rho_\nu/\bar{\rho_\nu}-1$. The position variance in each void pair (less than $100\ \rm{kpc}/h$) is negligible compared to
this scale and therefore void catalogue from either TianNu and TianZero gives same estimation of $\delta_{\nu}$.  
We choose this environment variable as proxy of the neutrino environment of the voids and the surrounding halos.
Larger (smaller)
$\delta_\nu$ implies that the halos of the region experience a stronger (weaker) neutrino free streaming and are more (less) influenced by the late time gravity interplay between massive neutrinos and CDM.

\section{results}\label{sec:results}

\subsection{Radius Variation}\label{subsec:Radius variation}
Based on the void catalogs of TianNu and TianZero simulations and the environment neutrino density of each void, we are then able to study the effect of massive neutrinos on the halo positions, reflected on the changes of the void radius. 
To this end, we measure the radius change between the TianNu and TianZero simulations, defined as $\Delta r \equiv r_\nu-r_0$, where $r_\nu$ and $r_0$ of each void pair are the radius obtained from TianNu and TianZero respectively. 
In Figure \ref{r_e_v} we first plot the distribution of voids as a function of the neutrino density contrast $\delta_\nu$ and radius $r$, shown in the upper panels, and then we rescale each pixel by the expectation value of the radius change in each pixel, and get the figure in the lower panels. The three columns show the results of the three different halo number densities which we choose to construct voids.    
From the upper panels, we can see the anti-correlation between $\delta_\nu$ and $r$, which indictes that voids of small (large) radius have generally high (low) $\delta_\nu$. 
The lower panels show that the voids are divided into two clusters in the $\delta _\nu$-$r$ plane, demonstrating that smaller voids are generally larger (the bluer cluster) in TianNu compared to the TianZero baseline while larger voids are generally smaller (the redder cluster).
These changes reveal the local changes in the halo positions, i.e., by the presence of massive neutrinos, halos surrounding the smaller voids are generally washed outward the void centers due to a stronger neutrino free streeming; halos surrounding the larger voids are generally squeezed toward the void centers due to a weaker neutrino free streeming.
This demonstrates the local effect of massive neutrinos on the spatial distribution of halos: the halo postions are locally changed according to different neutrino environment, toward a more uniform distribution in the massive neutrino cosmology.  

\subsection{Correlation Function}\label{sec:corr}
\begin{figure}[ht!]    
\center{\includegraphics[width=9cm]  {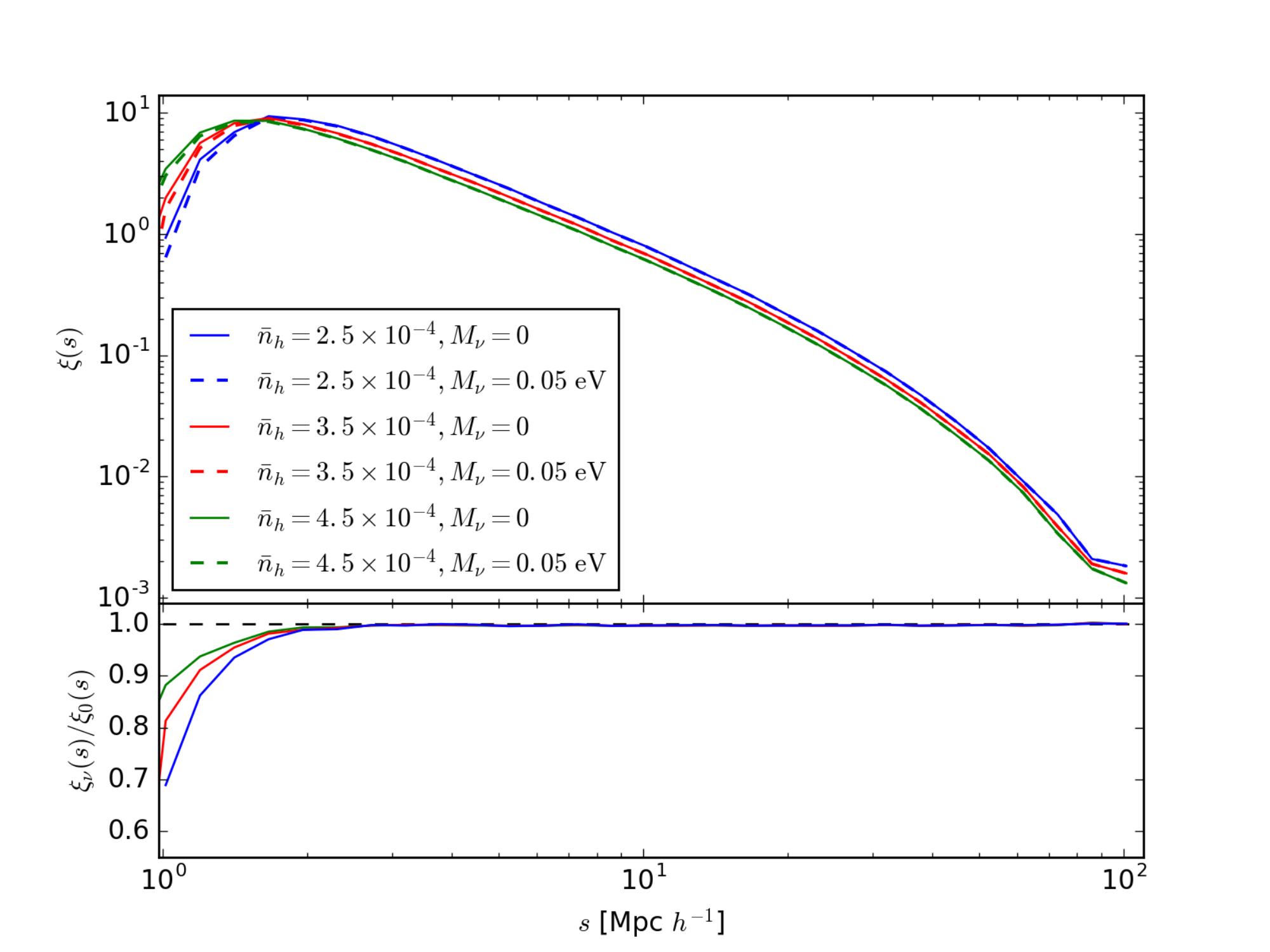}}        
\caption{Two-point correlation function of the TianNu (dashed lines) and TianZero (dotted lines) halos. The lower panel shows the ratios of TianNu to TianZero. The line colors represent different halo number densities $\bar{n}_h$ (upper and lower panels are consistent), in units of $h^3\rm{Mpc}^{-3}$. \label{corr}}
\end{figure}

This halo spatial distribution change can be detected by the 2-point correlation function (2PCF). Using the publicly available parallel code \textsc{CUTE} \footnote{http://members.ift.uam-csic.es/dmonge/CUTE.html} \citep{2012arXiv1210.1833A}, we plot in Figure \ref{corr} the halo 2PCFs of TianNu (dashed lines) and TianZero (solid lines), which are measured from the
most massive $\sim$0.42, $\sim$0.59 and $\sim$0.75 million halo pairs ($\bar{n}_h=[2.5,3.5,4.5]\times 10^{-4} $Mpc$^{-3}h^3$). The lower panel shows the ratios of TianNu to TianZero, with the three line colors corresponding to the three halo number densities $\bar{n}_h$. 
We can see that on scales smaller than 2 Mpc$/h$, the 2PCFs is slightly suppressed by neutrino mass, which reflects the late time gravity interply between massive neutrinos and the CDM halos.
On scales larger than 2 Mpc$/h$, the impact of neutrinos is negligible.

\subsection{Number Function}\label{subsec:Number function}
\begin{figure*}
\begin{minipage}[c]{0.49\linewidth}
\centering
\includegraphics[width=8cm]{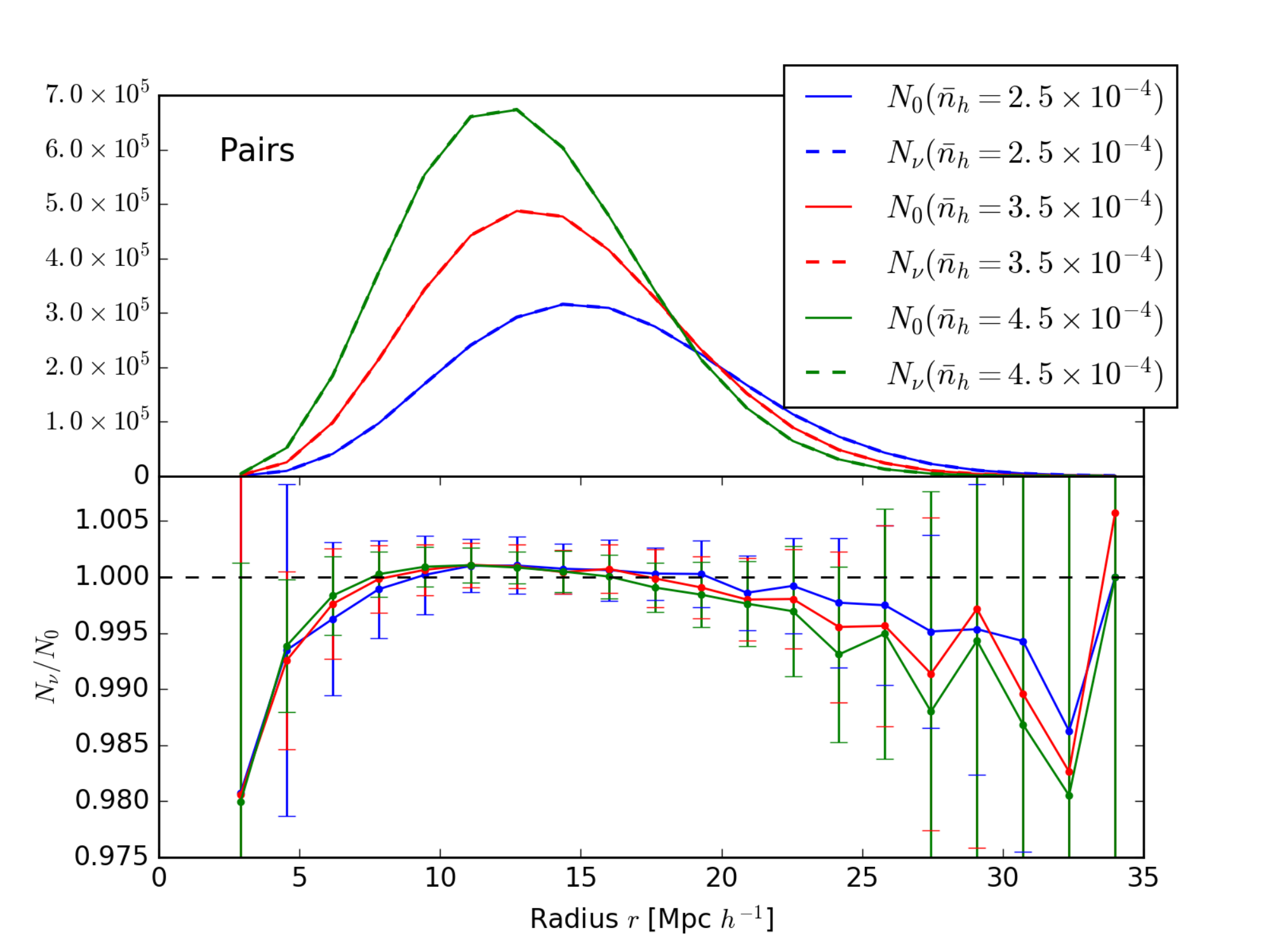}
\end{minipage}
\begin{minipage}[c]{0.49\linewidth}
\centering
\includegraphics[width=8cm]{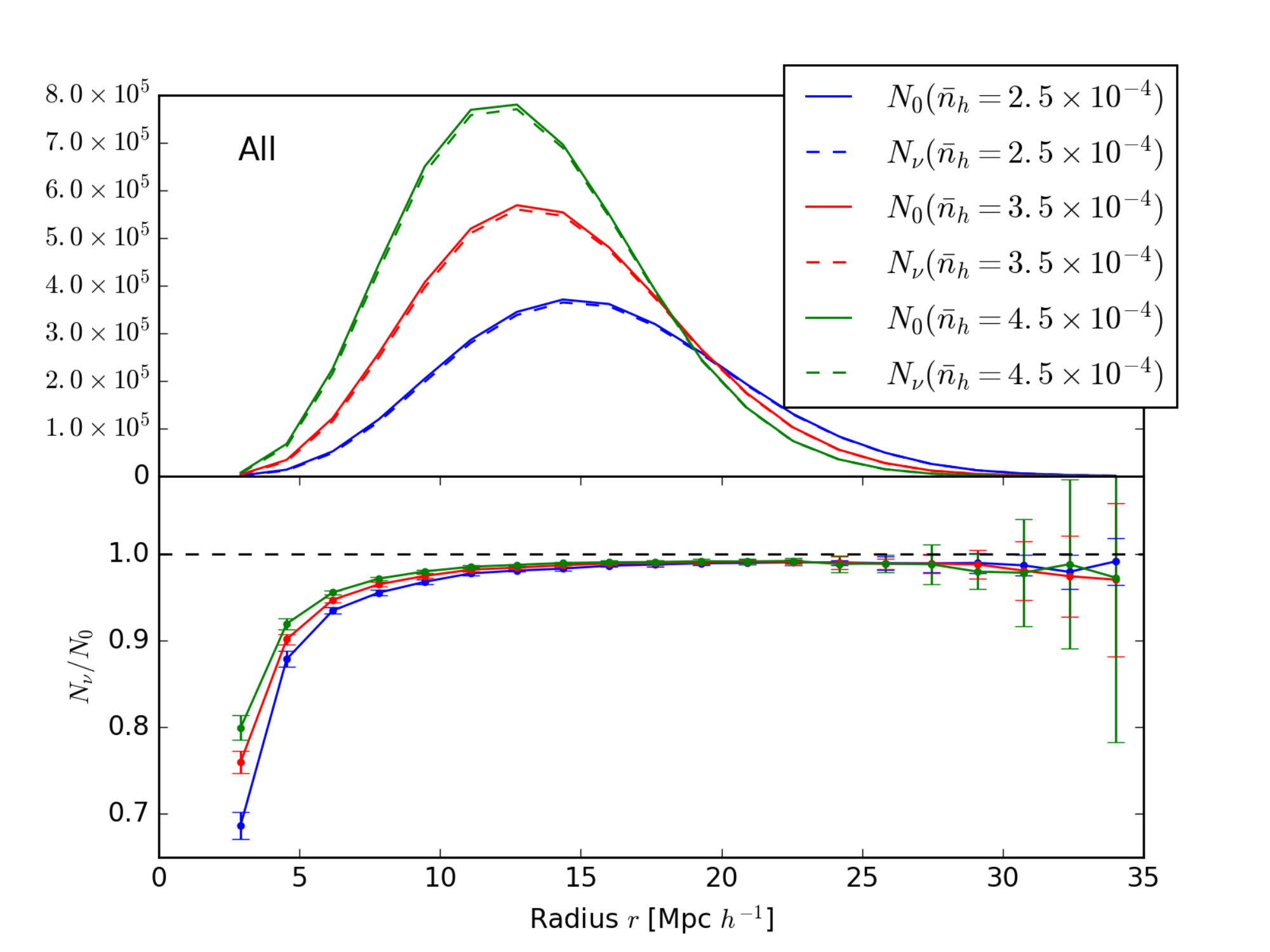}
\end{minipage}
\caption{Number function of voids pairs (left) and all voids (right) constructed from the TianNu and TianZero halos. The lower panels show the ratios of TianNu ($N_\nu$) to TianZero ($N_0$). The line colors represent different halo number densities $\bar{n}_h$ (upper and lower panels are consistent), in units of $h^3\rm{Mpc}^{-3}$. The error bars are from bootstrap resampling, which are too tiny to be visible in the upper panels. 
\label{r_func} }     
\end{figure*}

The void radius variances shown in Figure \ref{r_e_v} indicate that for TianNu compared to TianZero there are fewer big and small voids and more medium-sized voids. This can be reflected on the void number functions.   
Generally the number function is the void abundance as a function of their radius measured from the entire void catalog. However in order to compare the results shown in Figure \ref{r_e_v}, we first compute the number function from part of the void catalogs (the void pairs), shown in the left column of Figure \ref{r_func}.  
The lower panel shows the ratios of TianNu to TianZero with the line colors representing the void catalogs constructed from halos of different halo number densities and the error bars are measured from bootstrap resampling.   

We can see that the void abundances are suppressed in both the small and big radius ranges and are enriched in the intermediate ranges. However the difference between TianNu and TianZero is not comparable to the resampling noises, resulting from the fact that the changes of the halo positions caused by neutrinos is very small.
Taking into account the entire void catalogs, we plot in the right column of Figure \ref{r_func} the number functions computed from all the voids.  We find that for TianNu compared to TianZero the void abundance is smaller in almost all radius ranges and the number differences between TianNu and TianZero are about an order of magnitude larger than that computed from the void pairs. 
So the DT void number function demonstrates a good sensitivity for the halo position changes, which can potentially serve as supplement to the 2-point function to constrain the neutrino mass in current and future galaxy surveys.

\section{Summary}\label{sec:Summary}

Through the properties of DT void, we have studied in this work the local effect of massive neutrinos on the spatial distribution of CDM halos. To this end, we have applied the DIVE DT void finder to construct voids from the halo catalogs of TianNu and TianZero simulations. This has permitted us to find the local changes of the halo positions in regions of different neutrino environment, reflected on the changes of the void radius. In addition, we have plotted the 2-point correlation functions of the TianNu and TianZero halos and found the small scale suppression.
Finally, we have compared between TianNu and TianZero the number function of the void pairs and of all the voids and seek the void abundance differences.\par

Our study based on the void catalogs and the neutrino density field demonstrates that massive neutrinos impact the spatial distribution of the halos by locally changing the positions of the halos according to their neutrino environments.
The smaller voids are generally in higher neutrino density regions and halos surrounding the smaller DT voids are washed outward the void centers due to a stronger neutrino free streeming while halos surrounding the larger DT voids are squeezed toward the void centers.
The halos are therefore distributed more uniformly in the massive neutrinos cosmology, which suppresses the 2-point function of the halos on scales $\sim$1 Mpc$/h$.   
This spatial distribution change of the halos caused by neutrinos also skews the void number function, searving as a measurable neutrino effect and as supplement to the 2-point function to constrain the neutrino mass in current and future galaxy surveys. 

\section*{Acknowledgements}
We thank Shuo Yuan of Beijing university for his discussion and comment.
This work was supported by the National Science Foundation of China (Grants No. 11528306,11573006), the Fundamental ResearchFunds for the Central Universities, the Special Program for Applied Research on Super Computation of the NSFC-Guangdong Joint Fund (the second phase), and National Key R\&D Program of China (2017YFA0402600). 

\bibliographystyle{aasjournal}
\bibliography{mybib}

\begin{thebibliography}{}
\expandafter\ifx\csname natexlab\endcsname\relax\def\natexlab#1{#1}\fi
\providecommand{\url}[1]{\href{#1}{#1}}

\bibitem[{{Ahmad} {et~al.}(2001){Ahmad}, {Allen}, {Andersen}, {Anglin},
  {B{\"u}hler}, {Barton}, {Beier}, {Bercovitch}, {Bigu}, {Biller}, {Black},
  {Blevis}, {Boardman}, {Boger}, {Bonvin}, {Boulay}, {Bowler}, {Bowles},
  {Brice}, {Browne}, {Bullard}, {Burritt}, {Cameron}, {Cameron}, {Chan},
  {Chen}, {Chen}, {Chen}, {Chon}, {Cleveland}, {Clifford}, {Cowan}, {Cowen},
  {Cox}, {Dai}, {Dai}, {Dalnoki-Veress}, {Davidson}, {Doe}, {Doucas},
  {Dragowsky}, {Duba}, {Duncan}, {Dunmore}, {Earle}, {Elliott}, {Evans},
  {Ewan}, {Farine}, {Fergani}, {Ferraris}, {Ford}, {Fowler}, {Frame}, {Frank},
  {Frati}, {Germani}, {Gil}, {Goldschmidt}, {Grant}, {Hahn}, {Hallin},
  {Hallman}, {Hamer}, {Hamian}, {Haq}, {Hargrove}, {Harvey}, {Hazama},
  {Heaton}, {Heeger}, {Heintzelman}, {Heise}, {Helmer}, {Hepburn}, {Heron},
  {Hewett}, {Hime}, {Howe}, {Hykawy}, {Isaac}, {Jagam}, {Jelley}, {Jillings},
  {Jonkmans}, {Karn}, {Keener}, {Kirch}, {Klein}, {Knox}, {Komar}, {Kouzes},
  {Kutter}, {Kyba}, {Law}, {Lawson}, {Lay}, {Lee}, {Lesko}, {Leslie}, {Levine},
  {Locke}, {Lowry}, {Luoma}, {Lyon}, {Majerus}, {Mak}, {Marino}, {McCauley},
  {McDonald}, {McDonald}, {McFarlane}, {McGregor}, {McLatchie}, {Drees}, {Mes},
  {Mifflin}, {Miller}, {Milton}, {Moffat}, {Moorhead}, {Nally}, {Neubauer},
  {Newcomer}, {Ng}, {Noble}, {Norman}, {Novikov}, {O'Neill}, {Okada},
  {Ollerhead}, {Omori}, {Orrell}, {Oser}, {Poon}, {Radcliffe}, {Roberge},
  {Robertson}, {Robertson}, {Rowley}, {Rusu}, {Saettler}, {Schaffer},
  {Schuelke}, {Schwendener}, {Seifert}, {Shatkay}, {Simpson}, {Sinclair},
  {Skensved}, {Smith}, {Smith}, {Starinsky}, {Steiger}, {Stokstad}, {Storey},
  {Sur}, {Tafirout}, {Tagg}, {Tanner}, {Taplin}, {Thorman}, {Thornewell},
  {Trent}, {Tserkovnyak}, {van Berg}, {van de Water}, {Virtue}, {Waltham},
  {Wang}, {Wark}, {West}, {Wilhelmy}, {Wilkerson}, {Wilson}, {Wittich},
  {Wouters}, \& {Yeh}}]{2001PhRvL..87g1301A}
{Ahmad}, Q.~R., {Allen}, R.~C., {Andersen}, T.~C., {et~al.} 2001, Physical
  Review Letters, 87, 071301

\bibitem[{{Ahmad} {et~al.}(2002){Ahmad}, {Allen}, {Andersen}, {Anglin},
  {Barton}, {Beier}, {Bercovitch}, {Bigu}, {Biller}, {Black}, {Blevis},
  {Boardman}, {Boger}, {Bonvin}, {Boulay}, {Bowler}, {Bowles}, {Brice},
  {Browne}, {Bullard}, {B{\"u}hler}, {Cameron}, {Chan}, {Chen}, {Chen}, {Chen},
  {Cleveland}, {Clifford}, {Cowan}, {Cowen}, {Cox}, {Dai}, {Dalnoki-Veress},
  {Davidson}, {Doe}, {Doucas}, {Dragowsky}, {Duba}, {Duncan}, {Dunford},
  {Dunmore}, {Earle}, {Elliott}, {Evans}, {Ewan}, {Farine}, {Fergani},
  {Ferraris}, {Ford}, {Formaggio}, {Fowler}, {Frame}, {Frank}, {Frati},
  {Gagnon}, {Germani}, {Gil}, {Graham}, {Grant}, {Hahn}, {Hallin}, {Hallman},
  {Hamer}, {Hamian}, {Handler}, {Haq}, {Hargrove}, {Harvey}, {Hazama},
  {Heeger}, {Heintzelman}, {Heise}, {Helmer}, {Hepburn}, {Heron}, {Hewett},
  {Hime}, {Howe}, {Hykawy}, {Isaac}, {Jagam}, {Jelley}, {Jillings}, {Jonkmans},
  {Kazkaz}, {Keener}, {Klein}, {Knox}, {Komar}, {Kouzes}, {Kutter}, {Kyba},
  {Law}, {Lawson}, {Lay}, {Lee}, {Lesko}, {Leslie}, {Levine}, {Locke}, {Luoma},
  {Lyon}, {Majerus}, {Mak}, {Maneira}, {Manor}, {Marino}, {McCauley},
  {McDonald}, {McDonald}, {McFarlane}, {McGregor}, {Meijer Drees}, {Mifflin},
  {Miller}, {Milton}, {Moffat}, {Moorhead}, {Nally}, {Neubauer}, {Newcomer},
  {Ng}, {Noble}, {Norman}, {Novikov}, {O'Neill}, {Okada}, {Ollerhead}, {Omori},
  {Orrell}, {Oser}, {Poon}, {Radcliffe}, {Roberge}, {Robertson}, {Robertson},
  {Rosendahl}, {Rowley}, {Rusu}, {Saettler}, {Schaffer}, {Schwendener},
  {Sch{\"u}lke}, {Seifert}, {Shatkay}, {Simpson}, {Sims}, {Sinclair},
  {Skensved}, {Smith}, {Smith}, {Spreitzer}, {Starinsky}, {Steiger},
  {Stokstad}, {Stonehill}, {Storey}, {Sur}, {Tafirout}, {Tagg}, {Tanner},
  {Taplin}, {Thorman}, {Thornewell}, {Trent}, {Tserkovnyak}, {van Berg}, {van
  de Water}, {Virtue}, {Waltham}, {Wang}, {Wark}, {West}, {Wilhelmy},
  {Wilkerson}, {Wilson}, {Wittich}, {Wouters}, \& {Yeh}}]{2002PhRvL..89a1301A}
---. 2002, Physical Review Letters, 89, 011301

\bibitem[{{Alam} {et~al.}(2017){Alam}, {Ata}, {Bailey}, {Beutler}, {Bizyaev},
  {Blazek}, {Bolton}, {Brownstein}, {Burden}, {Chuang}, {Comparat}, {Cuesta},
  {Dawson}, {Eisenstein}, {Escoffier}, {Gil-Mar{\'{\i}}n}, {Grieb}, {Hand},
  {Ho}, {Kinemuchi}, {Kirkby}, {Kitaura}, {Malanushenko}, {Malanushenko},
  {Maraston}, {McBride}, {Nichol}, {Olmstead}, {Oravetz}, {Padmanabhan},
  {Palanque-Delabrouille}, {Pan}, {Pellejero-Ibanez}, {Percival}, {Petitjean},
  {Prada}, {Price-Whelan}, {Reid}, {Rodr{\'{\i}}guez-Torres}, {Roe}, {Ross},
  {Ross}, {Rossi}, {Rubi{\~n}o-Mart{\'{\i}}n}, {Saito}, {Salazar-Albornoz},
  {Samushia}, {S{\'a}nchez}, {Satpathy}, {Schlegel}, {Schneider},
  {Sc{\'o}ccola}, {Seo}, {Sheldon}, {Simmons}, {Slosar}, {Strauss}, {Swanson},
  {Thomas}, {Tinker}, {Tojeiro}, {Maga{\~n}a}, {Vazquez}, {Verde}, {Wake},
  {Wang}, {Weinberg}, {White}, {Wood-Vasey}, {Y{\`e}che}, {Zehavi}, {Zhai}, \&
  {Zhao}}]{2017MNRAS.470.2617A}
{Alam}, S., {Ata}, M., {Bailey}, S., {et~al.} 2017, \mnras, 470, 2617

\bibitem[{{Alonso}(2012)}]{2012arXiv1210.1833A}
{Alonso}, D. 2012, ArXiv e-prints, arXiv:1210.1833

\bibitem[{{Brandbyge} {et~al.}(2010){Brandbyge}, {Hannestad}, {Haugb{\o}lle},
  \& {Wong}}]{2010JCAP...09..014B}
{Brandbyge}, J., {Hannestad}, S., {Haugb{\o}lle}, T., \& {Wong}, Y.~Y.~Y. 2010,
  \jcap, 9, 014

\bibitem[{{Castorina} {et~al.}(2014){Castorina}, {Sefusatti}, {Sheth},
  {Villaescusa-Navarro}, \& {Viel}}]{2014JCAP...02..049C}
{Castorina}, E., {Sefusatti}, E., {Sheth}, R.~K., {Villaescusa-Navarro}, F., \&
  {Viel}, M. 2014, \jcap, 2, 049

\bibitem[{{Dawson} {et~al.}(2016){Dawson}, {Kneib}, {Percival}, {Alam},
  {Albareti}, {Anderson}, {Armengaud}, {Aubourg}, {Bailey}, {Bautista},
  {Berlind}, {Bershady}, {Beutler}, {Bizyaev}, {Blanton}, {Blomqvist},
  {Bolton}, {Bovy}, {Brandt}, {Brinkmann}, {Brownstein}, {Burtin}, {Busca},
  {Cai}, {Chuang}, {Clerc}, {Comparat}, {Cope}, {Croft}, {Cruz-Gonzalez}, {da
  Costa}, {Cousinou}, {Darling}, {de la Macorra}, {de la Torre}, {Delubac}, {du
  Mas des Bourboux}, {Dwelly}, {Ealet}, {Eisenstein}, {Eracleous}, {Escoffier},
  {Fan}, {Finoguenov}, {Font-Ribera}, {Frinchaboy}, {Gaulme}, {Georgakakis},
  {Green}, {Guo}, {Guy}, {Ho}, {Holder}, {Huehnerhoff}, {Hutchinson}, {Jing},
  {Jullo}, {Kamble}, {Kinemuchi}, {Kirkby}, {Kitaura}, {Klaene}, {Laher},
  {Lang}, {Laurent}, {Le Goff}, {Li}, {Liang}, {Lima}, {Lin}, {Lin}, {Lin},
  {Long}, {Lundgren}, {MacDonald}, {Geimba Maia}, {Malanushenko},
  {Malanushenko}, {Mariappan}, {McBride}, {McGreer}, {M{\'e}nard}, {Merloni},
  {Meza}, {Montero-Dorta}, {Muna}, {Myers}, {Nandra}, {Naugle}, {Newman},
  {Noterdaeme}, {Nugent}, {Ogando}, {Olmstead}, {Oravetz}, {Oravetz},
  {Padmanabhan}, {Palanque-Delabrouille}, {Pan}, {Parejko}, {P{\^a}ris},
  {Peacock}, {Petitjean}, {Pieri}, {Pisani}, {Prada}, {Prakash}, {Raichoor},
  {Reid}, {Rich}, {Ridl}, {Rodriguez-Torres}, {Carnero Rosell}, {Ross},
  {Rossi}, {Ruan}, {Salvato}, {Sayres}, {Schneider}, {Schlegel}, {Seljak},
  {Seo}, {Sesar}, {Shandera}, {Shu}, {Slosar}, {Sobreira}, {Streblyanska},
  {Suzuki}, {Taylor}, {Tao}, {Tinker}, {Tojeiro}, {Vargas-Maga{\~n}a}, {Wang},
  {Weaver}, {Weinberg}, {White}, {Wood-Vasey}, {Yeche}, {Zhai}, {Zhao}, {Zhao},
  {Zheng}, {Ben Zhu}, \& {Zou}}]{2016AJ....151...44D}
{Dawson}, K.~S., {Kneib}, J.-P., {Percival}, W.~J., {et~al.} 2016, \aj, 151, 44

\bibitem[{{Emberson} {et~al.}(2017){Emberson}, {Yu}, {Inman}, {Zhang}, {Pen},
  {Harnois-D{\'e}raps}, {Yuan}, {Teng}, {Zhu}, {Chen}, \&
  {Xing}}]{2017RAA....17...85E}
{Emberson}, J.~D., {Yu}, H.-R., {Inman}, D., {et~al.} 2017, Research in
  Astronomy and Astrophysics, 17, 085

\bibitem[{{Fukuda} {et~al.}(1998){Fukuda}, {Hayakawa}, {Ichihara}, {Inoue},
  {Ishihara}, {Ishino}, {Itow}, {Kajita}, {Kameda}, {Kasuga}, {Kobayashi},
  {Kobayashi}, {Koshio}, {Miura}, {Nakahata}, {Nakayama}, {Okada}, {Okumura},
  {Sakurai}, {Shiozawa}, {Suzuki}, {Takeuchi}, {Totsuka}, {Yamada}, {Earl},
  {Habig}, {Kearns}, {Messier}, {Scholberg}, {Stone}, {Sulak}, {Walter},
  {Goldhaber}, {Barszczxak}, {Casper}, {Gajewski}, {Halverson}, {Hsu}, {Kropp},
  {Price}, {Reines}, {Smy}, {Sobel}, {Vagins}, {Ganezer}, {Keig}, {Ellsworth},
  {Tasaka}, {Flanagan}, {Kibayashi}, {Learned}, {Matsuno}, {Stenger},
  {Takemori}, {Ishii}, {Kanzaki}, {Kobayashi}, {Mine}, {Nakamura}, {Nishikawa},
  {Oyama}, {Sakai}, {Sakuda}, {Sasaki}, {Echigo}, {Kohama}, {Suzuki}, {Haines},
  {Blaufuss}, {Kim}, {Sanford}, {Svoboda}, {Chen}, {Conner}, {Goodman},
  {Sullivan}, {Hill}, {Jung}, {Martens}, {Mauger}, {McGrew}, {Sharkey},
  {Viren}, {Yanagisawa}, {Doki}, {Miyano}, {Okazawa}, {Saji}, {Takahata},
  {Nagashima}, {Takita}, {Yamaguchi}, {Yoshida}, {Kim}, {Etoh}, {Fujita},
  {Hasegawa}, {Hasegawa}, {Hatakeyama}, {Iwamoto}, {Koga}, {Maruyama}, {Ogawa},
  {Shirai}, {Suzuki}, {Tsushima}, {Koshiba}, {Nemoto}, {Nishijima}, {Futagami},
  {Hayato}, {Kanaya}, {Kaneyuki}, {Watanabe}, {Kielczewska}, {Doyle}, {George},
  {Stachyra}, {Wai}, {Wilkes}, \& {Young}}]{1998PhRvL..81.1562F}
{Fukuda}, Y., {Hayakawa}, T., {Ichihara}, E., {et~al.} 1998, Physical Review
  Letters, 81, 1562

\bibitem[{{Inman} {et~al.}(2016){Inman}, {Yu}, {Zhu}, {Emberson}, {Pen},
  {Zhang}, {Yuan}, {Chen}, \& {Xing}}]{2016arXiv161009354I}
{Inman}, D., {Yu}, H.-R., {Zhu}, H.-M., {et~al.} 2016, ArXiv e-prints,
  arXiv:1610.09354

\bibitem[{{Kitaura} {et~al.}(2016){Kitaura}, {Chuang}, {Liang}, {Zhao}, {Tao},
  {Rodr{\'{\i}}guez-Torres}, {Eisenstein}, {Gil-Mar{\'{\i}}n}, {Kneib},
  {McBride}, {Percival}, {Ross}, {S{\'a}nchez}, {Tinker}, {Tojeiro},
  {Vargas-Magana}, \& {Zhao}}]{2016PhRvL.116q1301K}
{Kitaura}, F.-S., {Chuang}, C.-H., {Liang}, Y., {et~al.} 2016, Physical Review
  Letters, 116, 171301

\bibitem[{{Laureijs} {et~al.}(2011){Laureijs}, {Amiaux}, {Arduini},
  {Augu{\`e}res}, {Brinchmann}, {Cole}, {Cropper}, {Dabin}, {Duvet}, {Ealet},
  \& et~al.}]{2011arXiv1110.3193L}
{Laureijs}, R., {Amiaux}, J., {Arduini}, S., {et~al.} 2011, ArXiv e-prints,
  arXiv:1110.3193

\bibitem[{{Lesgourgues} {et~al.}(2013){Lesgourgues}, {Mangano}, {Miele}, \&
  {Pastor}}]{2013neco.book.....L}
{Lesgourgues}, J., {Mangano}, G., {Miele}, G., \& {Pastor}, S. 2013, {Neutrino
  Cosmology}

\bibitem[{{Lesgourgues} \& {Pastor}(2006)}]{2006PhR...429..307L}
{Lesgourgues}, J., \& {Pastor}, S. 2006, \physrep, 429, 307

\bibitem[{{Liang} {et~al.}(2016){Liang}, {Zhao}, {Chuang}, {Kitaura}, \&
  {Tao}}]{2016MNRAS.459.4020L}
{Liang}, Y., {Zhao}, C., {Chuang}, C.-H., {Kitaura}, F.-S., \& {Tao}, C. 2016,
  \mnras, 459, 4020

\bibitem[{{LSST Science Collaboration} {et~al.}(2009){LSST Science
  Collaboration}, {Abell}, {Allison}, {Anderson}, {Andrew}, {Angel}, {Armus},
  {Arnett}, {Asztalos}, {Axelrod}, \& et~al.}]{2009arXiv0912.0201L}
{LSST Science Collaboration}, {Abell}, P.~A., {Allison}, J., {et~al.} 2009,
  ArXiv e-prints, arXiv:0912.0201

\bibitem[{{Massara} {et~al.}(2015){Massara}, {Villaescusa-Navarro}, {Viel}, \&
  {Sutter}}]{2015JCAP...11..018M}
{Massara}, E., {Villaescusa-Navarro}, F., {Viel}, M., \& {Sutter}, P.~M. 2015,
  \jcap, 11, 018

\bibitem[{{Olive} \& {Particle Data Group}(2014)}]{2014ChPhC..38i0001O}
{Olive}, K.~A., \& {Particle Data Group}. 2014, Chinese Physics C, 38, 090001

\bibitem[{{Planck Collaboration} {et~al.}(2016){Planck Collaboration}, {Ade},
  {Aghanim}, {Arnaud}, {Ashdown}, {Aumont}, {Baccigalupi}, {Banday},
  {Barreiro}, {Bartlett}, \& et~al.}]{2016A&A...594A..13P}
{Planck Collaboration}, {Ade}, P.~A.~R., {Aghanim}, N., {et~al.} 2016, \aap,
  594, A13

\bibitem[{{Yu} {et~al.}(2017){Yu}, {Emberson}, {Inman}, {Zhang}, {Pen},
  {Harnois-D{\'e}raps}, {Yuan}, {Teng}, {Zhu}, {Chen}, {Xing}, {Du}, {Zhang},
  {Lu}, \& {Liao}}]{2017NatAs...1E.143Y}
{Yu}, H.-R., {Emberson}, J.~D., {Inman}, D., {et~al.} 2017, Nature Astronomy,
  1, 0143

\bibitem[{{Zhao} {et~al.}(2016){Zhao}, {Tao}, {Liang}, {Kitaura}, \&
  {Chuang}}]{2016MNRAS.459.2670Z}
{Zhao}, C., {Tao}, C., {Liang}, Y., {Kitaura}, F.-S., \& {Chuang}, C.-H. 2016,
  \mnras, 459, 2670

\end{thebibliography}
\end{document}